\documentclass[preprint]{emulateapj}
\usepackage{threeparttable}
\usepackage[backref,breaklinks,colorlinks,citecolor=blue]{hyperref}
\newcommand{\source}{PSR~J1640--4631}
\newcommand{\degree}{^{\circ}}

\begin{document}

    \author{R.~F. Archibald\footnotemark[1],
     ~E.~V. Gotthelf\footnotemark[2],
     ~R.~D. Ferdman\footnotemark[1],
     ~V.~M. Kaspi\footnotemark[1],
     ~S. Guillot\footnotemark[3],
     ~F.~A. Harrison\footnotemark[4],
     ~E.~F. Keane\footnotemark[5],
     ~M.~J. Pivovaroff\footnotemark[6],
     ~D.~Stern\footnotemark[7],
     ~S.~P. Tendulkar\footnotemark[1],
     ~and J.~A. Tomsick\footnotemark[8]}

     \footnotetext[1]{Department of Physics and McGill Space Institute, McGill University, 3600 University St., Montr\'eal QC, H3A 2T8, Canada}
     \footnotetext[2]{Columbia Astrophysics Laboratory, 550 West 120th Street, New York, NY, 10027-6601 USA}
     \footnotetext[3]{Instituto de Astrof\'{i}sica, Pontificia Universidad Cat\'{o}lica de Chile, Av. Vicu\~{n}a Mackenna 4860, 782-0436 Macul, Santiago, Chile}
     \footnotetext[4]{Cahill Center for Astrophysics, 1216 East California Boulevard, California Institute of Technology, Pasadena, California 91125, USA}
     \footnotetext[5]{SKA Organisation, Jodrell Bank Observatory, Cheshire, SK11 9DL, UK}
     \footnotetext[6]{Lawrence Livermore National Laboratory, 7000 East Ave. Livermore, CA 94550-9234 USA}
    \footnotetext[7]{Jet Propulsion Laboratory, California Institute of Technology, Pasadena, California 91109, USA}
    \footnotetext[8]{Space Science Laboratory, University of California, Berkeley, 7 Gauss Way Berkeley, CA 94720-7450, USA}

\title{A High Braking Index for a Pulsar}
\begin{abstract}
We present a phase-coherent timing solution for \source{}, a young 206\,ms pulsar using X-ray timing observations taken with  {\it NuSTAR}.
Over this timing campaign, we have measured the braking index of \source{} to be $n=3.15\pm0.03$.
Using a series of simulations, we argue that this unusually high braking index is not due to timing noise, but is intrinsic to the pulsar's spin-down.
We cannot, however, rule out contamination due to an unseen glitch recovery, although the recovery timescale would have to be longer than most yet observed.
If this braking index is eventually proven to be stable, it demonstrates that pulsar braking indices greater than 3 are allowed in nature, hence other physical mechanisms such as mass or magnetic quadrupoles are important in pulsar spin-down.
We also present a $3\sigma$ upper limit on the pulsed flux at 1.4\,GHz of  0.018 mJy.

\end{abstract}
\section{Introduction}
Pulsars emit light by extracting energy from their rotational kinetic stores.
As such, their spin-down is expected to follow the form \citep{1985Natur.313..374M}
\begin{equation}
    \label{eqn:bi}
    \dot{\nu}=-K\nu^n,
\end{equation}
where $\nu$ is the spin frequency of the pulsar, $\dot{\nu}$ the frequency derivative and $K$ a constant of proportionality related to the pulsar's moment of inertia and magnetic field structure \citep{1969Natur.221..454G}.
Here, $n$ is the braking index.
In the standard pulsar model of an unchanging magnetic dipole in a vacuum, electrodynamics predicts a value of three \citep[e.g.][]{1977puls.book.....M}.
In more realistic models of a pulsar and its magnetosphere, the braking index is predicted to always lie between 1.8 and 3 \citep{1997MNRAS.288.1049M}.
Values less than this can be obtained by relaxing the various assumptions of the model -- e.g. allowing magnetic field evolution \citep{1988MNRAS.234P..57B}, momentum loss due to a particle wind \citep{1999ApJ...525L.125H}, or a varying angle between the spin and magnetic poles \citep{2013Sci...342..598L}.

Taking the time derivative of Equation~\ref{eqn:bi} gives us the following fundamental equation which contains only the braking index and observable quantities:
\begin{equation}
    \label{eqn:n_obs}
    n=\frac{\nu\ddot{\nu}}{\dot{\nu}^2},
\end{equation}
where $\ddot{\nu}$ is the second-derivative of the spin frequency.
A measurement of $n$ can be made only for the youngest pulsars for which $\ddot{\nu}$ is large enough to detect on human timescales.
As such, only eight pulsars of the $\sim2400\;$known have measured braking indices, with values ranging from $0.9\pm0.2$ to $2.839\pm0.001$ \citep[][and references therein]{2015MNRAS.446..857L}.

\source{} was discovered as a pulsating X-ray source in a {\it NuSTAR} survey of the Norma region of the Galactic plane \citep{2014ApJ...788..155G}.
The pulsar is located in the center of the supernova remnant G338.1$-$0.0, and powers the pulsar wind nebula (PWN) HESS J1640$-$465, first detected in very high-energy gamma-rays  and thought to be the most luminous TeV source in our Galaxy \citep{2009ApJ...706.1269L}.
We undertook X-ray timing observations of \source{} starting shortly after discovery with the aim of measuring its braking index.
\section{Observations and Analysis}
\begin{figure}
    \centering
    \includegraphics[width=\columnwidth]{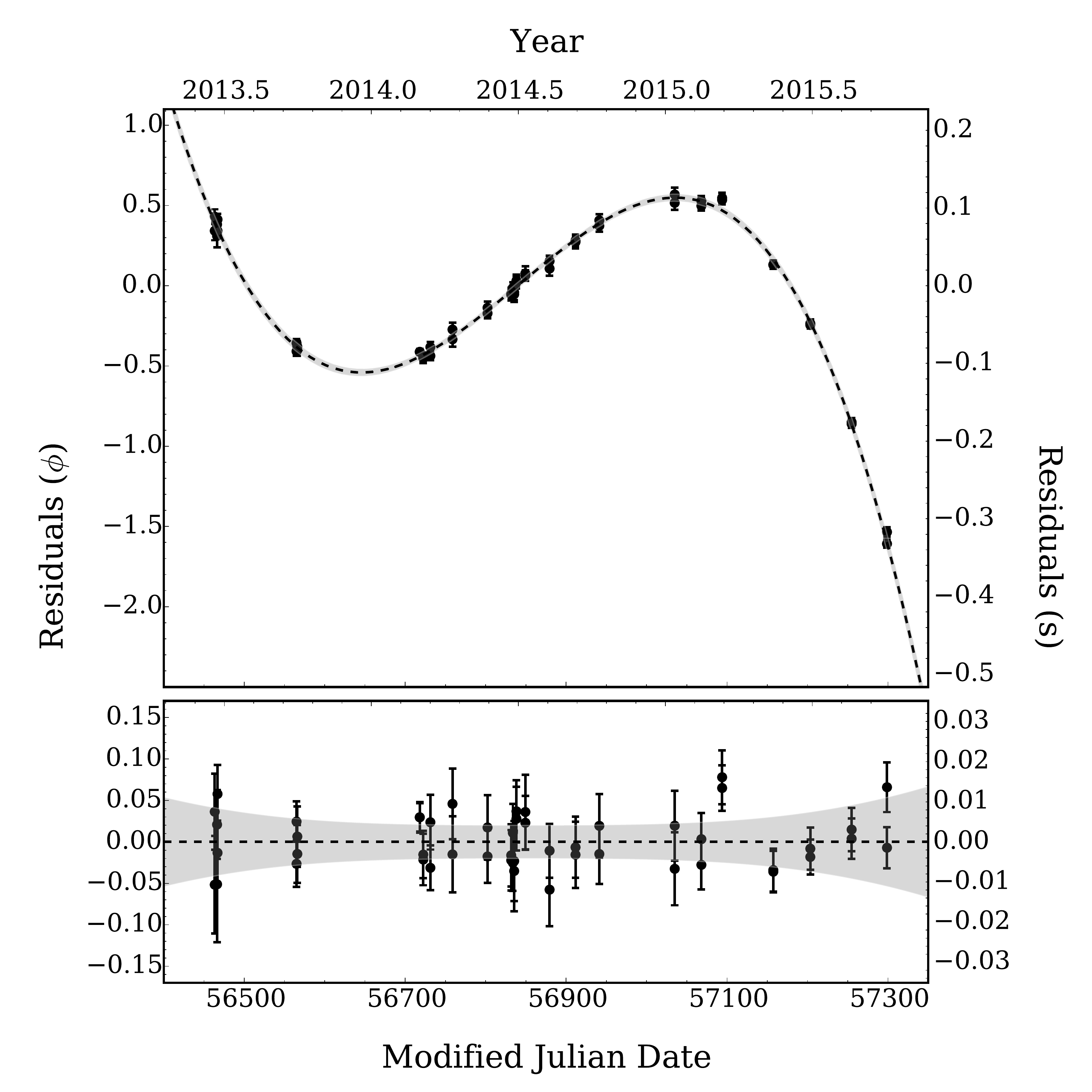}
    \caption[Residuals]{Timing residuals of \source{} from MJD 56463 to 57298,  29 September 2013 to 3 October 2015,  for the solution presented in Table~\ref{tab:timing}. The top panel shows the timing residuals subtracting only the contributions from $\nu$ and $\dot{\nu}$ with the dashed black line showing the fitted $\ddot{\nu}$ of (3.38$\pm$0.03)$\times10^{-22}$ s$^{-3}.$ The bottom panel shows the residuals after accounting for $\ddot{\nu}$. The gray bands in both panels indicate the 1-$\sigma$ timing model uncertainties.}
    \label{fig:resplot}
\end{figure}

All X-ray observations presented in this work were taken using {\it NuSTAR}, which consists of two co-aligned X-ray telescopes sensitive to photons with energies from 3--79$\;$keV \citep{2013ApJ...770..103H}.
{\it NuSTAR} observations of \source{} were typically 20--50$\;$ks, and the observation cadence can be seen in Figure~\ref{fig:resplot}.
Level 1 data products were obtained from {\tt HEASARC}  and reduced using {\tt nupipeline v0.4.4}.
 Photons from a circular region having a 30$''$ radius  centered on the source were extracted.
 To maximize the signal-to-noise ratio of the pulse, we used only photons with energies in the 3.0--55$\;$keV range.

Photon arrival times were corrected to the Solar System barycenter using the {\it Chandra} position of \source{}, RA= $16^{h}40^{m}43.52^{s}$ DEC=$-46^\circ31'35.4''$ \citep{2009ApJ...706.1269L} using {\tt barycorr} from {\tt HEASOFT v6.17} and {\tt v052} of the {\it NuSTAR} clock file.

Photon arrival times were used to derive an average pulse time-of-arrival (TOA) for each observation.
The TOAs were extracted using a Maximum Likelihood (ML) method.
The ML method compares a continuous model template of the pulse profile to the computed phases of the photon arrival times from an observation \citep{2009LivingstoneTiming}.
To create the template, all observations were folded into a high signal-to-noise profile.
This high signal-to-noise profile was then fitted to a Fourier model using the first two harmonics.
Two harmonics were chosen to optimally describe the pulse shape, as determined by use of the H-test. \citep{1989A&A...221..180D}.
 We verified that TOAs extracted using a cross-correlation method give consistent results.
  {\it NuSTAR}'s absolute timing calibration is accurate to $\pm3\;$ms \citep{2015arXiv150401672M}, smaller than our measurement uncertainties.

The TOAs were fitted to a standard timing model in which the phase as a function of time $t$ is described by a Taylor expansion:
\begin{equation}
    \label{eqn:phase}
    \phi(t) = \phi_0+\nu_0(t-t_0)+\frac{1}{2}\dot{\nu_0}(t-t_0)^2+\frac{1}{6}\ddot{\nu_0}(t-t_0)^3+\ldots
\end{equation}
This was done using the {\tt TEMPO2} \citep{2006MNRAS.369..655H} pulsar timing software package.
In Table~\ref{tab:timing} we present a fully phase-coherent timing solution for \source{} over the {\it NuSTAR} observation campaign.
This is the only solution which provides a statistically acceptable fit​, i.e. we have verified there are no pulse counting ambiguities.
The residuals, the difference between our timing model and the observed pulse phases,  can be seen in Figure~\ref{fig:resplot}.
In the top panel, we show these residuals accounting only for $\nu$ and $\dot{\nu}$.
The bottom panel shows the residuals for the full timing solution, accounting for $\ddot{\nu}$.
Fitting for an extra frequency derivative, $\dddot{\nu}$ does not significantly improve the fit with the F-test indicating a 52\% probability of the improvement of $\chi^2$ being due to chance.
We measure $\ddot{\nu}=(3.38\pm0.03)\times 10^{-22}\;$s$^{-3}$  corresponding to a braking index of $n=3.15\pm0.03$, where the uncertainty represents the 68\% confidence interval.

\begin{figure}
    \centering
    \includegraphics[width=\columnwidth]{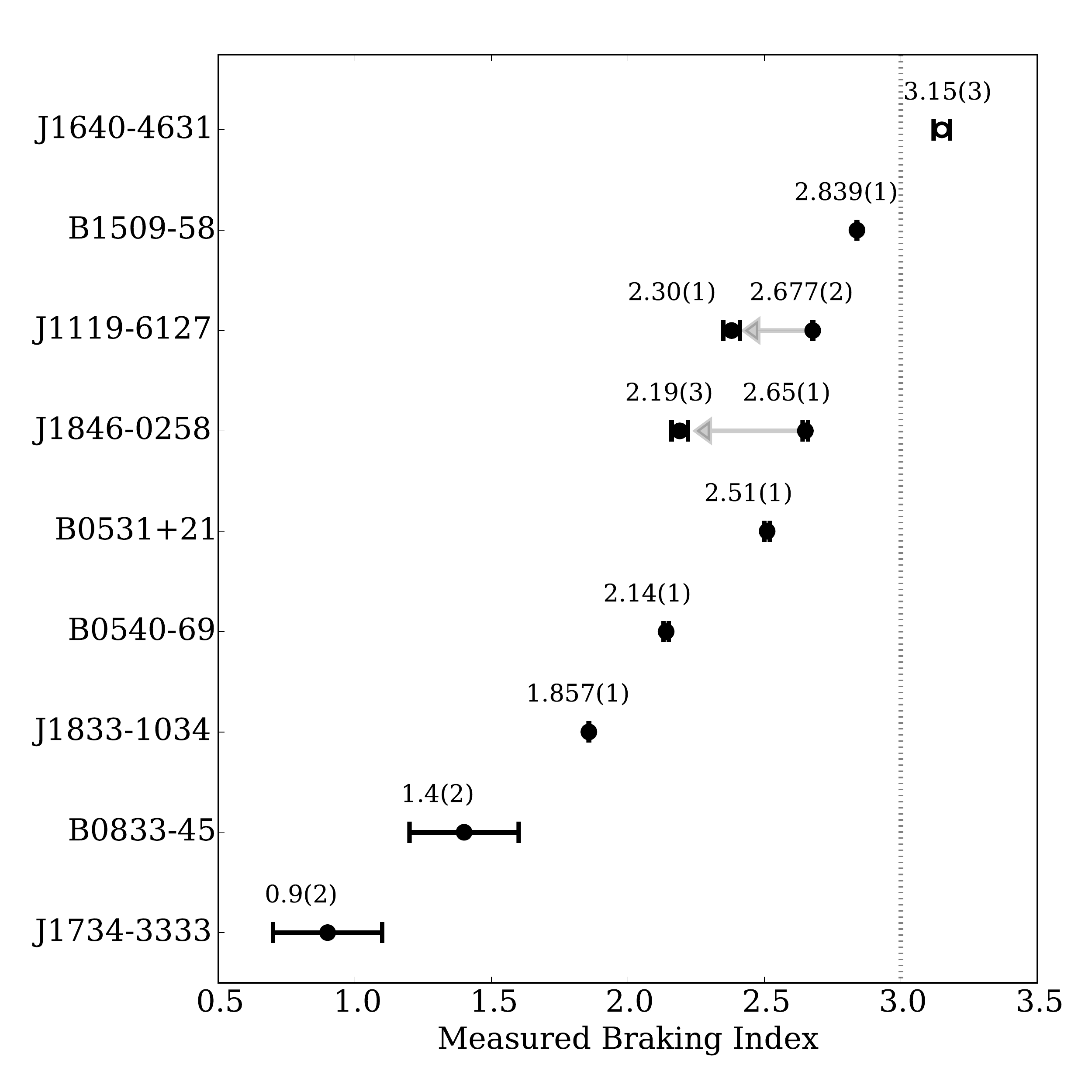}
    \caption[All measured braking indices.]{All braking indices where the long-term electrodynamic spin-down is believed to be dominant.The gray dotted line indicates a braking index of three, that which is expected for a pure magnetic dipole. For PSR~J1846$-$0258 \citep{2015ApJ...810...67A} and PSR~J1119$-$6127 \citep{2015MNRAS.447.3924A}, where the braking index changed following glitches, the gray arrows indicate the direction of change following the glitch. All other braking indices are from \cite{2015MNRAS.446..857L} and references therein.}
    \label{fig:all_ns}
\end{figure}

This measured braking index is $5\sigma$ higher than that expected in the standard magnetic dipole scenario.
In Figure~\ref{fig:all_ns}, we show all braking indices where the long-term electrodynamic spin-down is believed to be dominant; note how \source{} is the only measurement greater than the canonical $n=3$ magnetic dipole line.

\begin{table}
    \begin{center}
    \caption{Phase-Coherent Timing Parameters for \source{}.}
    \label{tab:timing}
    \begin{tabular}{ll}
    \hline
    \hline
    Dates (MJD)         & 56463.0--57298.8\\
    Dates               & 29 September 2013 -- 3 October 2015 \\
    Epoch (MJD)         & 56741.00000\\
    $\nu\;$ (s$^{-1}$)            & 4.843 410 287 0(5)\\
    $\dot{\nu}\;$ (s$^{-2}$)            &$-$2.280 830(4)$\times 10^{-11}$\\
    $\ddot{\nu}\;$ (s$^{-3}$)            & $3.38(3)\times 10^{-22}$ \\
    $|\dddot{\nu}|\;$(s$^{-4}$)      & $ < 1.4\times 10^{-30}$  \\
    rms residual (ms) & 6.17\\
    rms residual (phase) & 0.030\\
    $\chi^2_\nu$/dof & 0.98/46 \\
    Braking index, $n$                              & 3.15(3)\\
    \hline
    \hline
    \newline
    \end{tabular}
    \newline
    Note: Figures in parentheses are  the nominal 1$\sigma$ \textsc{tempo2} uncertainties in the least-significant digits quoted. Upper limits are quoted at the $2\sigma$ level. The source position was held fixed at the {\it Chandra} position.
    \end{center}
\end{table}

\subsection{Timing Noise Simulations}
A possible way to explain such a large measured braking index is contamination from timing noise.
Timing noise refers to unexplained low-frequency modulations found in the timing residuals of many, particularly young, pulsars \citep{1994ApJ...422..671A}.
Timing noise in radio pulsars has been observed to be spectrally `red', that is, having most power at low frequencies \citep{1994ApJ...422..671A}.
As such, it can contaminate measurements of $\ddot{\nu}$, and thus $n$ \citep{2010MNRAS.402.1027H}.
The power spectral density (PSD) of timing noise is often  modeled as
\begin{equation}
    \Phi_{TN}(f)=A\left(1+\frac{f^2}{f_c^2}\right)^{-q/2},
\end{equation}
where $A$ is the spectral density amplitude, $f_c$ the corner frequency, and $q$ the power-law index  \citep{2015MNRAS.449.3293L}.
A $q$ of 0 represents a white noise power spectrum, whereas
indices of 2, 4, and 6 represent random walks in pulse arrival phase, the pulse frequency, and frequency derivative, respectively.

To quantify the probability of timing noise biasing our measurement of the braking index, we conducted a series of simulations that aimed to determine whether any reasonable form of red noise artificially results in a measurement of $n>3$, given the observed (i.e. white) noise properties of the resulting timing residuals.
As timing noise in real pulsars can have PSDs with many indices we created $10^5$  realizations of red noise and injected them into simulated pulsar TOAs, with TOA uncertainties,  $\nu$, and $\dot{\nu}$ identical to that of the pulsar, but with a braking index of $n=3$.

To do this we used the {\tt simRedNoise} plug-in of {\tt TEMPO2}.
We simulated parameters on a grid, drawing 10 iterations from each set of parameters.
 We simulated $A$ ranging from {\bf $10^{-20}$--$10^{-18}\;$}s$^{2}$yr$^{-1}$ with 20 log-spaced steps, $q$ between 0--6 with 25 linear steps, and $f_c$ from $10^{-3}$--$10^{0.5}\;$ yr$^{-1}$ in 20 log-space steps.
The upper bound on $A$ was chosen to ensure phase connection was possible, as larger values of $A$ precluded phase connection more that 50\% of the time, and hence are ruled out by our observations.

After the noise injection, the new TOAs were fitted to the full timing model to measure $n$, allowing $\nu$, $\dot{\nu}$, and $\ddot{\nu}$ to vary.
We considered any iterations where the $\chi^2$ value indicated that a probability of less than 1\% ruled out by our measured residuals, as this would indicate unaccounted for noise.
After this,  only in $0.01\%$ of all the simulations could produce a braking index greater than three with the measured significance.
The parameter regime which gave this highest probability of artificially producing a high braking index was when $f_c$ was of order the observing length with the highest values of $A$.
There is only a weak dependence on $q$, with larger values producing more false positives.

Another way to check for the possible contamination of our measured value of $n$ by timing noise is by considering the third frequency derivative.
In our data, the third frequency derivative is consistent with zero at the $1\sigma$ level,
We ran another suite of simulations within the parameter space described above wherein we fitted up to a third frequency derivative.
We note that there is a covariance, with the second and third frequency derivatives being anti-correlated.
Only in $0.008\%$ of simulated sets of residuals can we reproduce a braking index significantly greater than three without having a detectable  third frequency derivative.

These simulations indicate that only a very low level of timing noise can be present in our data, and the measured braking index of $n=3.15\pm0.03$ is highly unlikely to be due to timing noise.
We note that the assumed value of $n=3$ in our simulations is conservative, since when assuming $n<3$, it is even less likely for timing noise to result in a measured $n>3$.

\subsection{Parkes Observations}
\label{sec:radio}
In order to search for radio pulsations from \source{}, we undertook observations with the 64-m Parkes Telescope.
Observations were performed in two sessions totaling 14.96 hours at the position of \source{} \citep{2009ApJ...706.1269L}.
Data were taken by observing with the central beam of the 21\,cm Multi-beam receiver, using the BPSR pulsar backend.
These observations were taken on 2014 April 18 and 2015 April 27 (MJDs 56765 and 56775, respectively) at center frequency 1382 MHz over 400 MHz of bandwidth, divided into 1024 channels.  The data from each channel were detected and the two polarizations summed to form a time series with 64-$\mu$s samples.

We searched these data using the known pulsar spin frequency and frequency derivative from the phase-coherent timing analysis found in this article.
We searched over 4704 dispersion measures between 0 and 1600\,pc\,cm$^{-3}$.  No signal was found in these data, and so we quote an upper limit to the pulsar flux at this frequency, using the radiometer equation for the rms noise from the observing system:
\begin{equation}
    \sigma_\textrm{rms} = \frac{T_{\textrm{sys}}}{G\sqrt{n_p t_{\textrm{obs}} \Delta f}},
\end{equation}
where $T_{\textrm{sys}}$ is the system temperature in Kelvin, $G$ is the receiver gain in K/Jy, $n_p$ is the number of polarizations, $t_{\textrm{obs}}$ is the total integration time in seconds, and $\Delta f$ is the observing bandwidth in Hz.
From this, we calculate a $3\sigma$ upper flux limit of 0.018 mJy at 1.4\,GHz.
This upper limit assumes a 50\% duty cycle, with the upper limit scaling as $\sqrt{DC/(1-DC)}$, where DC is the duty cycle.
This flux limit is low, but not unusually so, especially when one considers that the estimated distance to the source is $\sim$12~kpc.

\section{Discussion \& Conclusions}

A possible contaminant to the measured braking index is a long-term recovery from an unseen glitch prior to our monitoring.
If such a glitch occurred, a typical exponential recovery would, in general, cause an artificially high value for $\ddot{\nu}$ to be measured \citep[e.g.][]{1999MNRAS.306L..50J,2010MNRAS.402.1027H}.
Indeed, for pulsars with $\tau_c < 10^5 \;$yrs, there is a clear preference for positive $\ddot{\nu}$, compared to older pulsars which are equally likely to have positive or negative ${\ddot{\nu}}$ \cite[e.g.][]{2010MNRAS.402.1027H}.
If we assume an exponentially recovering glitch ​contaminating a constant braking index we can, from our upper limit of $\dddot{\nu}$, place a lower limit on the decay timescale for such a glitch to be $\tau=250\sqrt[3]{\Delta\nu_d/10^{-8}\mathrm{Hz}}$\,days, where $\nu_d$ is the size of the unseen decaying glitch.
In \cite{2013MNRAS.429..688Y}, of the 107 glitches detected, 27 had a detected exponential recovery. Of those, only 3 were longer than 200 days.
We thus cannot rule out contamination due to a prior unseen glitch recovery as an explanation for the high measured braking index, although the recovery time scale would have to be longer
than most yet observed.
Ultimately this hypothesis can be tested by continued monitoring.

Measurements of braking indices could also be contaminated by uncertainties in the pulsar's position, or proper motion \citep{1993Natur.366..663B}.
The position of \source{} is well determined by {\it Chandra} to a 3$\sigma$ error radius of 0.6'' \citep{2009ApJ...706.1269L}.
This positional error at the ecliptic latitude of \source{} would add a 1.0\,ms root-mean-square signal to our timing residuals, far smaller than our measurement uncertainties.
For \source{}'s estimated distance of 8--13$\;$kpc \citep{2009ApJ...706.1269L}, and a typical pulsar kick velocity of 300\,km\,s$^{-1}$ \citep{1997MNRAS.291..569H}, an unmodeled proper motion would change the measured braking index by less than one part in a million.
Thus neither of these effects can account for our measured braking index.

The pulsar's high luminosity PWN has been argued to be powered by a relativistic outflow or wind from the neutron star.
Under this assumption, predictions for the pulsar's spin-down history and hence braking index have been made \citep{2014ApJ...788..155G}.
The 0.2--10$\;$TeV luminosity of the PWN powered by the source represents $\sim6\%$ of the pulsar's current spin-down luminosity \citep{2014ApJ...788..155G}.
A pulsar whose spin-down is driven solely by a particle wind would result in a braking index of one \citep{1969ApJ...158..727M,1985Natur.313..374M}.
Furthermore, a combination of magnetic dipole radiation and wind braking would result in a braking index with value between one and three.
In this case, the braking index as a function of $\epsilon$, the fraction of spin-down power due to a particle wind, is given by \citep{2015MNRAS.446..857L}
\begin{equation}
    n=\frac{2}{\epsilon+1}+1.
\end{equation}
This implies a maximal expected braking index of $n$=2.89 for \source{}.
Indeed, a more thorough modeling of the pulsar and PWN system suggested an even smaller braking index, $n\approx 1.9$ \citep{2014ApJ...788..155G}, clearly at odds with our result.

A changing magnetic field has also been put forth as a possibility for a braking index that is different from three by the growth or decay of the field \citep{1988MNRAS.234P..57B, 2015MNRAS.446.1121G}, or a change in the angle between the magnetic and rotation axes\citep{2013Sci...342..598L}.
In this case, $n$ is given by:
\begin{equation}
    n=3+2\frac{\nu}{\dot{\nu}}\left(\frac{\dot{B}}{B}+\frac{\dot{\alpha}}{\mathrm{tan}\alpha}\right).
\end{equation}
For the decaying field case, this would imply a magnetic field decay rate of $\sim200\;$MG per century.
This decay rate is very close to that predicted in  some magneto-thermal evolutionary models \citep{2013MNRAS.434..123V}, and in this interpretation might be providing direct observational evidence of a decaying magnetic field.
On the other hand, population synthesis studies find no strong evidence for field decay in the radio pulsar population as a whole \citep{2006ApJ...643..332F}.

If a change in the alignment angle between the magnetic and rotational axes is the cause of the anomalous braking index then either $\alpha$ is less than $\pi/2$ and the rotation and magnetic axes are moving towards alignment, or $\alpha$ is greater than $\pi/2$ and the rotation and magnetic axes are counter-aligning.
If $\alpha$ is changing on order of the rate of the Crab pulsar \citep{2013Sci...342..598L}, the only pulsar for which such a change has been measured,  at $\sim1\degree{}$ per century, this would imply that $\alpha$ is $\sim5\degree{}$ away from being an orthogonal rotator.
This is at odds with the pulse profile being single peaked, since an orthogonal rotator would typically be seen to have emission as each pole enters our line of sight.
In general, the value of $\alpha$ for pulsars can be independently determined by modeling of the gamma-ray or radio pulse profiles.
\source, however, is radio quiet, see \S\ref{sec:radio} .
There are also no detected gamma-ray pulsations from \source{} \citep{2014ApJ...788..155G}, and so the value of $\alpha$ is unknown for this source.

Another possibility to explain a braking index greater than three is to invoke higher order multipoles \citep{2015MNRAS.450..714P}.
A pure quadrupole, either a magnetic quadrupole, or a mass quadrupole leading to gravitational radiation \citep{1988MNRAS.234P..57B},  would yield a braking index of 5, and could coexist with the magnetic dipole to give a braking index between 3 and 5.
Analogous to the case of a wind, the fraction of spin down due to a quadrupole versus a dipole, $\epsilon_Q$ is \citep{2000A&A...354..163P}:
\begin{equation}
    \epsilon_Q=\frac{n-3}{5-n}.
\end{equation}
In our case, this implies $\sim8\%$ of the spin down is due to quadrupolar radiation.
In the case of a mass quadrupole, this would imply that the pulsar has an ellipticity of $\sim0.005$, which cannot be reproduced by theoretically proposed dense matter equations of state, for a neutron star rotating at 4.84\,Hz \citep{2005PhRvL..95u1101O}.
 If such an ellipticity did exist, it would produce gravitational waves having a maximum strain of $\sim4\times 10^{-26}\left(\frac{12 {\rm kpc}}{d}\right)$ at twice the spin-period of the pulsar \citep{2000A&A...354..163P}, which is far below the detection sensitivity of current technology.

The existence of a magnetic quadrupole is in principle testable with future X-ray polarimeter missions.
X-ray polarization measurements of neutron stars are in principle sensitive to the magnetospheric configuration, \citep{2006MNRAS.373.1495V,2014MNRAS.438.1686T} be it a quadrupolar field structure or a change in the alignment of the spin and magnetic poles.
The specific magnetic field structure of pulsars has a strong impact on the inferred magnetic field strength, as well as predicted radio and gamma-ray pulse profiles \citep{2015MNRAS.450..714P}.
Thus X-ray polarimetric observations of \source{} could help us understand the origin of the pulsar's high $n$, and shed light on the range of possibilities of neutron-star magnetic field structure.

Since the first measurement of the Crab's braking index in 1972 \citep{1972ApJ...175..217B}, we have known that various physical mechanisms, such as angular momentum loss due to a wind, can result in a pulsar braking index less than the canonical dipole value.
Our results for \source{} now show that other physics, such as the quadrupole moment of the magnetic field, affect the evolution of this source, and likely rotation-powered pulsars, in general.
Given that two other young, high-magnetic field pulsars have experienced glitches that resulted in a significant drop in the braking index \citep{2015ApJ...810...67A,2015MNRAS.447.3924A}, it is clear that continuous study of braking indices provides an important window into additional physical processes at work in the youngest and most energetic of neutron stars.

\section*{Acknowledgements}
This work made use of data from the {\it NuSTAR} mission, a project led by the California Institute of Technology, managed by the Jet Propulsion Laboratory, and funded by the National Aeronautics and Space Administration.
Parkes radio telescope is part of the Australia Telescope National Facility which is funded by the Commonwealth of Australia for operation as a National Facility managed by CSIRO.
We also thank an anonymous referee for helpful comments that improved the manuscript.
R.F.A. acknowledges support from an  NSERC  Alexander Graham Bell Canada  Graduate Scholarship.
E.V.G. received support from the National Aeronautics and Space Administration through Chandra Award Number GO5-16061X issued by the Chandra X-ray Observatory Center, which is operated by the Smithsonian Astrophysical Observatory for and on behalf of the National Aeronautics Space Administration under contract NAS8-03060.
V.M.K. receives support from an NSERC Discovery Grant and Accelerator Supplement, Centre de Recherche en Astrophysique du Queb\'ec, an R. Howard Webster Foundation Fellowship from the Canadian Institute for Advanced Study, the Canada Research Chairs Program, and the Lorne Trottier Chair in Astrophysics and Cosmology.
Part of this work was performed under the auspices of the U.S. Department of Energy by Lawrence Livermore National Laboratory under Contract DE-AC52-07NA27344.

\end{document}